\documentclass[twocolumn,amsmath,amssymb]{revtex4}
\usepackage{graphicx}
\usepackage{setspace}
\usepackage{epsfig}
\begin{document}
\title{ Front propagation in A$\rightarrow$2A, A$\rightarrow$3A process
in $1d$: velocity, diffusion and velocity correlations. }

\author{Niraj Kumar and Goutam Tripathy } 
\address{Institute of Physics, Sachivalaya Marg, Bhubaneswar 751005,
India}

\begin{abstract}
 We study front propagation in the reaction
 diffusion process $\{A\stackrel{\epsilon}\rightarrow2A, A\stackrel
 {\epsilon_t}\rightarrow3A\}$ on a one dimensional ($1d$) lattice with 
 hard core interaction between the particles. Using the leading particle
 picture, velocity of the front in the system is computed using different 
 approximate methods, which is in good agreement
 with the simulation results. It is observed that in certain ranges of 
 parameters, the front velocity varies as a power law of $\epsilon_t$, which
 is well captured by our approximate schemes. We also observe that the front
 dynamics exhibits temporal velocity correlations and these must be 
 taken care of in order to find the exact estimates for the front diffusion 
 coefficient. This correlation changes sign depending upon the sign of 
 $\epsilon_t-D$, where $D$ is the bare diffusion coefficient of $A$ 
 particles. For $\epsilon_t=D$, the leading particle and thus the front 
 moves like an uncorrelated random walker, which is explained through
 an exact analysis.
\end{abstract}
\maketitle{}
\section{Introduction}
Front propagation is an important field of study in nonequilibrium 
systems. We often encounter these propagating fronts separating 
different phases in physics, chemistry and biology \cite{sar}. Here, in 
this work, we  study the dynamics of the front in the 
reaction-diffusion system $A\rightarrow2A$, $A\rightarrow3A$ 
in one dimensional lattice. At the macroscopic level, the mean field 
theory yields the following partial differential equation for the 
coarse grained concentration $\rho(x,t)$,
\begin{eqnarray}{\label{e1}}
\frac{\partial\rho}{\partial t}=D\frac{\partial^2\rho}{\partial x^2}+
   2\epsilon\rho(1-\rho)+\epsilon_t\rho(1-\rho)^2,
\end{eqnarray}
where, $D$ is the diffusion coefficient of the particle and 
$\epsilon$ and $\epsilon_t$ are the rates of single ($A\rightarrow2A$)
and twin ($A\rightarrow3A$) offspring 
production respectively. Equation (\ref{e1}) reduces to
the well known Fisher equation \cite{fisher} when $\epsilon_t=0$,
which models the reaction diffusion equation $A\rightarrow2A$. The
microscopic lattice model for $A\rightarrow2A$ has been studied 
extensively \cite{bram}\cite{ker1}\cite{ker2}\cite{ng}. The mean 
field allows travelling wave solution of the form $\rho(x,t)=\phi(x-vt)$,
where the velocity of an initially sharp front between $\rho=1$ ( stable ) 
and $\rho=0$ ( unstable ) state approaches an asymptotic velocity $V_0=2\sqrt{(
2\epsilon+\epsilon_t)D}$.
\section{Model, Front velocity and diffusion coefficient}
 We consider a $1d$ lattice composed of sites $i=1,2...L$.
 We start with the step function like distribution where, the
 left half is filled with $A$ particles while the right half is 
 empty. Each site $i$ can either be empty or occupied by maximum
 one particle i.e. hard core exclusion is enforced. We
 update the system random sequentially where $L$ microscopic moves
 correspond to one Monte Carlo step (MCS). During each updating we 
 randomly select a site and the particle at the site can undergo one
 of the following three microscopic moves.\\
 (1) The particle can jump to neighbouring empty site 
    with rate $D$,\\
 (2) The particle can give birth of one particle at either of the 
    empty neighbouring sites with rate $\epsilon$,\\
 (3) The particle can generate two new particles at 
    both the neighbouring sites provided both are empty with rate $\epsilon_t$.\\
    These processes are shown in the Fig. (\ref{fig:moves_1d}).\\
    As time evolves, these stochastic moves result in the stochastic
    movement of the front. As has been argued in \cite{ng}, the front may
    be identified with the rightmost $A$ particle.
    In this paper, we are interested in the dynamics of front whose
    evolution may be described by the following master equation\cite{van}.
 \begin{eqnarray}{\label{e4}}
 \frac{dP(X,t)}{dt}=(\epsilon+D)P(X-1,t)+\epsilon_tQ_0(X-1,t)+\nonumber\\
        DQ_0(X+1,t)-(\epsilon+D)P(X,t)-(\epsilon_t+D)Q_0(X,t)
 \end{eqnarray}
    Here, $P(X,t)$ is the probability distribution of finding the 
    front particle at position $X$ at a time $t$ and $Q_0(X,t)$ is
    the joint probability that the front is at $X$ at time $t$ {\it{and}} the
    site just behind it is empty. In Eq. (\ref{e4}), the
    first term corresponds to the forward hopping  of the front 
    particle from the position $X-1$ to reach  $X$ due to birth of
    a single particle or diffusion. The second term corresponds to  
    twin production at left and right neighbouring sites of site 
    $X-1$ and which results in the front moving from $X-1$
    to $X$. The third term corresponds to the backward hopping of 
    the front particle from position $X+1$ due to diffusion, provided
    site $X$ is empty. Last two terms account for the possible 
    jumps of the front particle from position $X$, which leads to front 
    moving either at $X-1$ or $X+1$. The dynamical properties of the 
    front that we want to study are its velocity $V$ and the diffusion
    coefficient $D_f$ which are defined as:
    \begin{eqnarray}{\label{e5}}
     V=\frac{d}{dt}<X(t)>
    \end{eqnarray} 
    \begin{eqnarray}{\label{e6}} 
     D_f=\frac{1}{2}\frac{d}{dt}<(X(t)-<X(t)>)^2> 
    \end{eqnarray}
    Where, $<X(t)>=\sum_XXP(X,t)$.
    Now we use the Eq. (\ref{e4}) and normalization $\sum P(X)=1$ and
    taking $Q_0(X)=(1-\rho_{X-1})P(X)$, where $\rho_{X-1}$ is
    the probability that site $X-1$ is occupied. Denoting $\rho_{X-1}=\rho_1$,
    we obtain the following expression for the asymptotic velocity and 
    diffusion coefficient of the front.
    \begin{eqnarray}{\label{e7}}
     V=\epsilon+\epsilon_t-\rho_1(\epsilon_t-D)
    \end{eqnarray}
    \begin{eqnarray}{\label{e8}}
     D_f=\frac{1}{2}\{\epsilon+\epsilon_t+2D-\rho_1(\epsilon_t+D)\}
    \end{eqnarray}
    In order to obtain the velocity and diffusion coefficient we 
    need to know $\rho_1$, which is the density of site just 
    behind the front. In \cite{bram} for $\epsilon_t=0$, it was 
    shown that front velocity approaches asymptotically the mean 
    field value $V=V_0$ in the limit $\frac{D}{\epsilon}\rightarrow
   \infty$, while $V=\epsilon+D$ in the opposite limit $\frac{D}
   {\epsilon}\rightarrow 0$. But, when we are in between these two
   extreme limits we need to know $\rho_1$ and we expect similar features
   when $\epsilon_t\ne0$. There is no method to find 
   this value exactly. Here we present some approximate analytic estimates
   for $\rho_1$ and hence the front velocity. In subsection A, we use fixed
   site representation method, where a truncated master equation is written
   in the frame moving with the front, as discussed in \cite{ng}. In 
   subsection B, we apply two particle representation scheme proposed by
   Kerstein \cite{ker2} while in C a mixed scheme is proposed which yields
   better results than either A or B.
 \subsection{Fixed site representation}
\begin{figure}
     \centering
    \includegraphics[width=2.5in,height=1.50in]{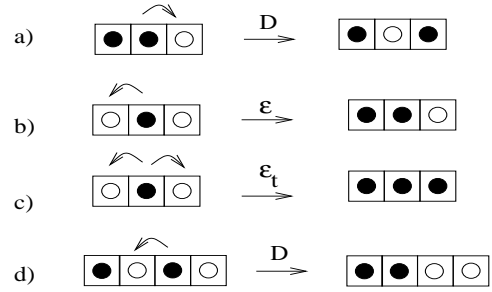}
    \caption{Microscopic moves, rightmost $\bullet$ represents the 
    front. (a) Diffusion of the front particle to its 
    right site leading to transition from $\bullet\bullet$ to 
    $\circ\bullet$ with rate $D$. (b) Creation of one particle to the 
    left of the front leads to transition from $\circ\bullet$ to
    $\bullet\bullet$ with rate $\epsilon$. (c) $\circ\bullet$ changes
    to  $\bullet\bullet$ due to creation of twins at both neighbouring
    sites of the front with rate $\epsilon_t$. (d) $\circ\bullet
    \rightarrow\bullet\bullet$, if the front takes diffusive move
    to its left and the second site behind the front is occupied. This
    occurs with rate $D\rho_2$, where $\rho_2$ is the probability of 
    occupancy at the second site behind the front.}
    \label{fig:moves_1d}
     \end{figure}

\begin{figure}
     \centering
    \includegraphics{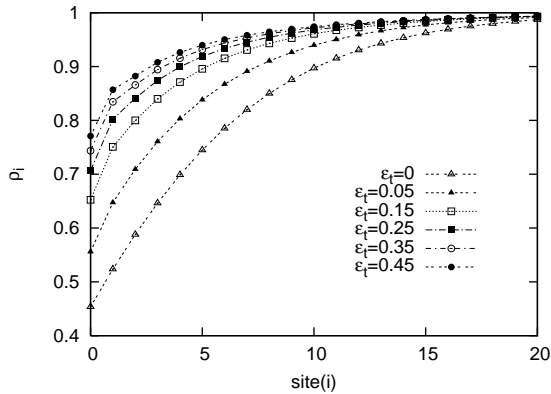}
    \caption{Density profile behind the front for different values of
    $\epsilon_t$ while keeping $D=0.25$ and $\epsilon=0.025$ fixed. Here, we
    notice that as $\epsilon_t$ decreases density profile curve shifts away
    from the bulk density level.}
    \label{fig:den_profile}
    \end{figure}

    This method has been proposed in \cite{ng} for the reaction 
    diffusion process $A\leftrightarrow2A$. Here, we write a truncated 
    master equation in the frame moving with the front. The simplest
    set of states is:$\{\circ\bullet, \bullet\bullet\}$, which 
    corresponds to the evolution of occupancy
    at a site just behind the front particle($l=1$). Here the rightmost
    $\bullet$ in each state corresponds to the front particle. These 
    two states make transitions between each other due to the 
    microscopic processes in the system as shown in 
    Fig. (\ref{fig:moves_1d}).
    Considering all such transitions the evolution of probabilities of
    these two states are given by:
    \begin{eqnarray}{\label{e9}}
    \frac{dP(\circ\bullet)}{dt}&=&(2D-D\rho_{2})P(\bullet\bullet)
          -\{2D\rho_{2}+\epsilon(2+\rho_{2})\nonumber\\
	  & &+\epsilon_t(1+\rho_{2}(1-\rho_{3}))\}P(\circ\bullet),\nonumber\\
    \frac{dP(\bullet\bullet)}{dt}&=&\{2D\rho_{2}+\epsilon(2+
                                   \rho_{2})+\epsilon_t
           (1+\rho_{2}(1-\rho_{3}))\nonumber\\& &\}P(\circ\bullet)
			   -(2D-D\rho_{2})P(\bullet\bullet).
   \end{eqnarray}
   Here, $\rho_i$ is the density at the $i$th site 
   behind the front and we have neglected the spatial density 
   correlation between consecutive pairs of sites beyond the second
   site behind the front. Now, using Eq. (\ref{e9}) and 
   normalization $P(\circ\bullet)+P(\bullet\bullet)=1$, we obtain
   the following expression for $\rho_1$:
   \begin{eqnarray}{\label{e10}}
    \rho_1=\frac{2D\rho_{2}+\epsilon(2+\rho_{2})+\epsilon_t(1+
                     \rho_{2}(1-\rho_{3}))}
                    {D\rho_{2}+2D+2\epsilon+\epsilon\rho_{2}+
                    \epsilon_t(1+\rho_{2}(1-\rho_{3}))}
   \end{eqnarray}
   From Eq. (\ref{e10}), we note that in order to find $\rho_1$ we 
   need to know $\rho_2$ and $\rho_3$. As a crude approximation if we 
   assume that $\rho_{2}=\rho_{3}=\rho^{b}=1$, where $\rho^{b}$ is the bulk 
   density, we get the following value of $\rho_1$.
   \begin{eqnarray}{\label{e11}}
   \rho_1\simeq\frac{2D+3\epsilon+\epsilon_t}{3D+3\epsilon+\epsilon_t}
   \end{eqnarray}
   Now using this approximation for $\rho_1$ in Eq. (\ref{e7}), we find the
   estimate for the velocity which is in reasonable agreement with
   the simulation, as shown in the Fig. (\ref{fig:velocity})
   . The estimate for $V$ can be improved by including more sites
   in the truncated representation. For example, for $l=2$ we study the 
   evolution of following set of four states: $\{\circ\circ\bullet, \circ\bullet
   \bullet, \bullet\circ\bullet, \bullet\bullet\bullet\}$ and as expected
   we get improved results as shown in Fig. (\ref{fig:velocity}). Here, we 
   notice that for larger values of $\epsilon_t$ the simulation results 
   show nice
   agreement with that of analytic results. However, as $\epsilon_t$
    decreases and approaches  zero, we see gradual departure of the
    simulation data from the analytic one. In fact, $\rho_i$ differs
    from the bulk density significantly with decreasing value $\epsilon_t$ as
    shown in the Fig. (\ref{fig:den_profile}). That is, the 
    approximation $\rho_i\approx1$ holds better for larger 
    values of $\epsilon_t$ and hence we get better agreement with 
    the simulation results. The estimate for the velocity
    can be further improved if we include states with larger number 
    of sites. In  Fig. (\ref{fig:velocity}) we notice 
    two interesting points: firstly, for $D=\epsilon_t$, the theoretical
    result matches strikingly with the simulation result, secondly, we observe
    a power law dependence of the velocity on $\epsilon_t$. In fact, the first
    point, can be shown to be exact by noting that when $D=\epsilon_t$,
    the front velocity
    from Eq.(\ref{e7}) is $V=\epsilon+\epsilon_t$, which
    is independent of $\rho_1$.
   \subsection{Two particle representation}
\begin{figure} 
  \centering
  \includegraphics[width=2.2in,height=1.8in]{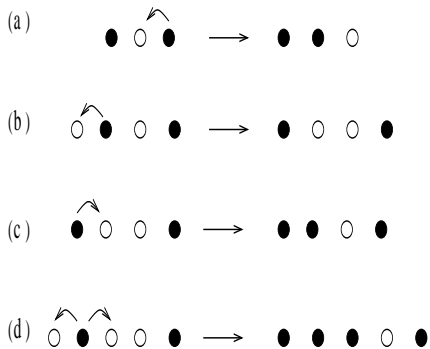}
  \caption{Transition between two particle states with rightmost
      $\bullet$ representing front.
     (a) Diffusive move of the front particle to its 
     left leading to transition 101$\rightarrow$11 with
     rate D, (b) When the second particle behind the
     front jumps to the left, provided it is empty, state
     changes from 101$\rightarrow$1001 with rate $D(1-\rho)$,
     (c) Birth of a single particle by the second particle
     to its left with rate $\epsilon$ leads to transition
     1001$\rightarrow$101, (d) 1001$\rightarrow$101 if the
     second particle gives birth of two particles, provided the
     the site left to it is empty, with rate $\epsilon_t(1-\rho)$.}
   \label{fig:moves_2P}
  \end{figure}

   In the following, we try to find the analytic
   estimates for $\rho_1$ using Kerstein's two particles
   representation\cite{ker2}. In this representation each state of 
   the system is defined by two rightmost particles and thus
   we have an infinite set of states: $\{11, 101, 1001,
   10001, 100001,....\}$. Here, the rightmost '1' represents
   the front particle while the leftmost '1' is the second
   particle behind it and '0' stands for empty site. Let us denote
   by $P_k$ the probability of two particle state with $k$ empty
   sites between the leading particle and next particle behind it. These
   states form a closed set under transition due to microscopic processes. We
   have illustrated few transitions in the Fig. (\ref{fig:moves_2P}). 
   Considering all such transitions and denoting the probability of
   occupancy of site just behind the second particle by $\rho$, we
   write the following rate equations for $P_k$.   
  \begin{eqnarray}{\label{e13}}
   \frac{dP_0}{dt}&=&(\epsilon+2D)P_1+\epsilon_t(1-\rho)P_1
     +(2\epsilon+\epsilon_t)(1-P_0)\nonumber\\
     &  &-(2D-D\rho)P_0,\nonumber\\
   \frac{dP_k}{dt}&=& (2D-D\rho)P_{k-1}+\{\epsilon+2D+
   \epsilon_t(1-\rho)\}P_{k+1}\nonumber\\
   &-&(4D-D\rho+3\epsilon+2\epsilon_t-\epsilon_t\rho)P_k,
   {\hspace{0.4cm}}k\ge1. 
  \end{eqnarray}
    In order to solve Eq.(\ref{e13}) we need to 
    specify the dependence of $\rho$ on the parameters($\epsilon,
    \epsilon_t,D$). Following Kerstein \cite{ker2}, we write $\rho=aP_0-bP_0^2$ 
    and enforcing the condition that $\rho=1$ when $P_0=1$, we write the 
    following expression for $\rho$.
    \begin{eqnarray}{\label{e14}}
    \rho=(1+\lambda)P_0-\lambda P_0^2
    \end{eqnarray}
    This equation specifies the dependence of $\rho$ on the parameters
    implicitly through dependence of $P_0$ on $\epsilon,
    \epsilon_t,D$. Here, $\lambda$ is a free parameter to be evaluated
    as follows. Following Kerstein, using the ansatz $P_k=P_0(1-P_0)^k$ and 
    Eq. (\ref{e14})
    in Eq.(\ref{e13}), we get the following quartic equation 
    in $P_0$.
    \begin{eqnarray}{\label{e15}}
    \epsilon_t\lambda P_0^4&+&(D\lambda-\epsilon_t-2\epsilon_t
    \lambda)P_0^3+(\epsilon+D
    +2\epsilon_t+\epsilon_t\lambda\nonumber\\& & -D\lambda )P_0^2
    +\epsilon P_0-2\epsilon-\epsilon_t=0
    \end{eqnarray}
    In order to find $P_0$ we need to fix the value of $\lambda$. 
    For large $D$ and $\epsilon_t=0$, it is known that the front
    particle moves with its mean field velocity\cite{bram}. If we assume that
    this also happens when $\epsilon_t\neq 0$, then  equating the 
    mean field  front velocity $V_0=2\sqrt{(2\epsilon+\epsilon_t)}D$
    with that obtained from Eq. (\ref{e7}) i.e. $V\sim DP_0$ when $D$ 
    is very large compared to other parameters, we get
    $P_0=2\sqrt{\frac{2\epsilon+\epsilon_t}{D}}$. Using
    this value of $P_0$ in Eq. (\ref{e15}) we find $\lambda=3/4$
    in the limit $D\rightarrow\infty$. We solve the quartic equation 
    (\ref{e15}) to get the value of $\rho_1=P_0$ and hence the front 
    velocity, as shown in the Figs. (\ref{fig:velocity}) and 
    (\ref{fig:rho1_vel_et0}) and marked as $2P$. 
   
   \subsection{Mixed representation}
\begin{figure}
    \centering
   \includegraphics[width=2.2in,height=1.8in]{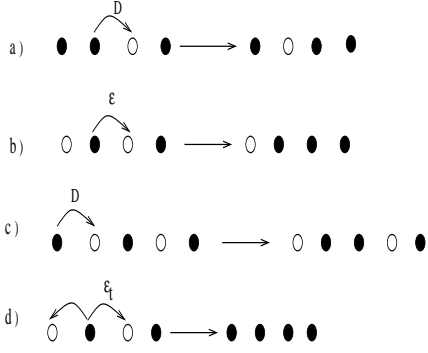}
   \caption{Transitions between mixed particle states with the 
   rightmost $\bullet$ representing front.
      (a) Diffusive move of the particle to the right empty site
           with rate D.This leads to transition from the state
	   (1,1) to (0,0).
      (b) Birth of a new particle on the right neighbouring empty
           site with rate $\epsilon$, which changes the state (1,0)
	   to (0,1).
      (c) Transition from (1,0) to (1,1) with rate D$\rho$, when the 
          third particle jumps to the right neighbouring empty site.
      (d) (1,0)$\rightarrow$(0,1) when the second particle behind the
      front in (1,0) realization produces twins at the neighbouring empty
      sites with rate $\epsilon_t$.
      }
   \label{fig:moves_2}   
   \end{figure}

    Since we are dealing with a multiparticle interacting system
   it is always desirable to include as many particles as possible
   while studying the evolution of the system. The simplest extension
   to the two particle representation is to study the evolution of the 
   following set of states: $\{\circ\bullet\bullet, \bullet\bullet
   \bullet , \circ\bullet\circ \bullet, \bullet\bullet\circ \bullet
   ...\}$, where the rightmost $\bullet$ in each state denotes the front
    particle. Since in this representation, each state is characterized by two 
   or three particles and hence we name it as mixed representation
    (MR). The rightmost $\bullet$ in each state denotes the front
    particle. When viewed in the frame moving with the front, each 
    state contains the location of second particle and the occupancy
    of the site just behind the second particle. We denote these states
    as $(k, 0)$ or $(k, 1)$, representing the states having $k$ empty sites
    between the front and the second particle and the site just after the 
    second particle is empty or occupied respectively. For example, by
    (0, 0) we mean the state $\circ\bullet\bullet$ and (0, 1) for the
    state $\bullet\bullet\bullet$. These states are making transitions
    among each other due to the microscopic processes and form a 
    closed set. We have shown some of the transitions in the Fig. 
    (\ref{fig:moves_2}).

   Assuming $\rho$ as the density of site, which is, next nearest
   neighbour to the second particle, we write the following
   rate equation  for the evolution of probabilities $P(k,0)$ and $P(k,1)$,
    $k=0,1,..\infty$.
   \begin{eqnarray}{\label{16}}
   \frac{dP(0,1)}{dt}&=&\{D\rho+\epsilon\rho+2\epsilon+\epsilon_t
        \rho(1-\rho)\}P(0,0)\nonumber\\&+&(D+2\epsilon+\epsilon_t)
	P(1,1)+(2\epsilon+2\epsilon_t)P(1,0)\nonumber\\&+&
	\epsilon_t\{P(2,0)+P(2,1)+P(3,0)+P(3,1)+...\}\nonumber\\
	&-&(2D-D\rho)P(0,1),\nonumber\\
   \frac{dP(0,0)}{dt}&=&D(1-\rho)P(0,1)+(D+\epsilon)P(1,1)\nonumber\\
        &+&(2D+\epsilon)P(1,0)+2\epsilon\{P(2,1)+P(2,0)\nonumber\\
	&+&P(3,0)+P(3,1)+...\}-\{2D+2\epsilon+D\rho\nonumber\\
	&+&\epsilon\rho+\epsilon_t\rho(1-\rho)\}P(0,0),\nonumber\\
   \frac{dP(k,1)}{dt}&=&DP(k-1,1)+D\rho P(k-1,0)\nonumber\\
         &+&\{D\rho+\epsilon+\epsilon\rho+\epsilon_t\rho(1-\rho)
	 \}P(k,0)\nonumber\\&+&(D+\epsilon)P(k+1,1)+(\epsilon+
	 \epsilon_t)P(k+1,0)\nonumber\\&-&(4D+3\epsilon-D\rho+
	 \epsilon_t)P(k,1),\nonumber\\
    \frac{dP(k,0)}{dt}&=&(D+D(1-\rho))P(k-1,0)+D(1-\rho)P(k,1)
         \nonumber\\&+&DP(k+1,1)+2DP(k+1,0)-\{4D+4\epsilon
	 \nonumber\\&+&D\rho+\epsilon\rho+2\epsilon_t+
	 \epsilon_t\rho(1-\rho)\}P(k,0).
   \end{eqnarray}
   In order to find $P_0$, we need to solve
   the above set of coupled equations. However, one can 
   find the analytic estimate for $P_0$ by solving rate equations
   for $P(0,0)$ and $P(0,1)$ and assuming $P(1,1)=\rho P_1, P(1,0)=
   (1-\rho)P_1$. Using $\displaystyle\sum_{i=0}^{1}P(k,i)=P_k$ 
   and $\displaystyle\sum_{k=0}^{\infty}P_k=1$, we find steady state expression
   for $P(0,0)$ and $P(0,1)$ in terms of $P_1$ and $\rho$ and then solve the 
   equation:
   \begin{eqnarray}{\label{e17}}
   P(0,0)+P(0,1)=P_0
   \end{eqnarray}
   Following Kerstein \cite{ker2}, if we use the ansatz $P_1=P_0(1-P_0)$, we get
   the following equation.
   \begin{eqnarray}{\label{e18}}
   \alpha P_0(1-P_0)+\beta P_0+\gamma=0
   \end{eqnarray}
    Where,
    \begin{eqnarray}{\label{e19}}
     \alpha &=&(2\epsilon+\epsilon_t+D\rho-\epsilon_t\rho)\{3D+2\epsilon
     +\epsilon\rho+\epsilon_t\rho(1-\rho)\}\nonumber\\& &+(2D-D\rho-\epsilon)\{2D
     +2\epsilon+\epsilon\rho+\epsilon_t\rho(1-\rho)\},\nonumber\\ 
     \beta &= & \{2D+2\epsilon+\epsilon\rho+D\rho+\epsilon_t\rho
     (1-\rho)\}(2D-D\rho)\nonumber\\& &-\{D\rho+2\epsilon+\epsilon\rho+\epsilon_t
     \rho(1-\rho)\}(D-D\rho),\nonumber\\
     \gamma &=& \epsilon_t(1-P_0)\{3D+2\epsilon+\epsilon\rho+\epsilon_t
     \rho(1-\rho)\}\nonumber\\& &+2\epsilon(1-P_0)\{2D+2\epsilon+\epsilon\rho
     +\epsilon_t\rho(1-\rho)\}.
    \end{eqnarray}
    
\begin{figure}
     \centering
    \includegraphics[bb=52 51 221 176 ]{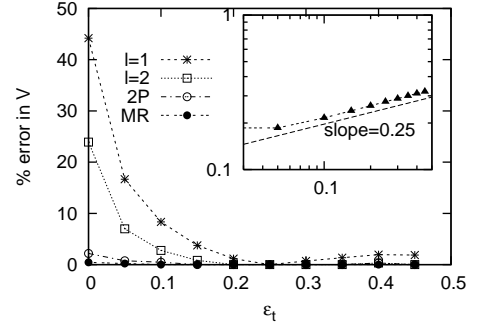}
    \caption{Percentage relative error in $V$, i.e. $\frac{|V^s-V^a|}{V^s}\times100$
    ($V^s$ and $V^a$ representing simulation and analytic results respectively),
    is plotted against $\epsilon_t$ while 
    keeping D=0.25 and $\epsilon=0.025$ fixed. Here stars, open squares, open 
    circles and filled circles correspond to the analytic estimates using 
    $l=2$(two states), $l=3$ (four states), two particle(2P)
    and mixed representation (MR) respectively. Simulation 
    profile and analytic profile using MR are essentially
    coincident and hence we notice almost zero error.
    Inset: log-log plot for the velocity versus 
    $\epsilon_t$, the straight line shows power law $V\approx
    \epsilon_t^{0.25}$.}
    \label{fig:velocity}
    \end{figure}

\begin{figure}
    \centering
   \includegraphics{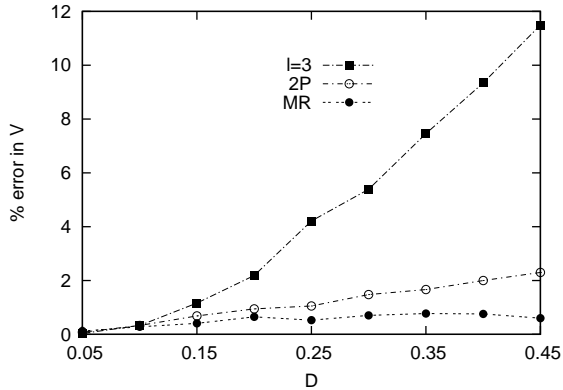}
   \caption{Percentage relative error in $V$ versus $D$, keeping 
      $\epsilon=0.05$ and $\epsilon_t=0$. The top data(filled squares) 
      corresponds to ref.\cite{ng}, where in the frame moving with the 
      front ,the evolution of particles at three sites behind
      the front was studied and assuming the fourth site at the 
      bulk density. The middle data (open circles) corresponds to 
      Kerstein \cite{ker2} two particle self consistent representation
      while the bottom data (filled circles) is the result from 
      the mixed representation.}
   \label{fig:rho1_vel_et0}   
   \end{figure}

   The Eq. (\ref{e18}) is in terms of two unknowns $\rho$ and 
   $P_0$ and hence we need to know $\rho$ in order to find $P_0$.
   We specify the dependence of $\rho$ on $P_0$ similar to what
   we did in the case of two particle representation. Using
   the results obtained for $P_0=\rho_1$ in Eq.(\ref{e7}) we find
   the estimates for $V$ as shown in 
   the Figs. (\ref{fig:velocity}) and 
    (\ref{fig:rho1_vel_et0}) marked as MR. We observe good 
   agreement with the simulation results.
    
   \section{Front diffusion coefficient}
\begin{figure}
     \centering
    \includegraphics[width=3.0in,height=2.5in]{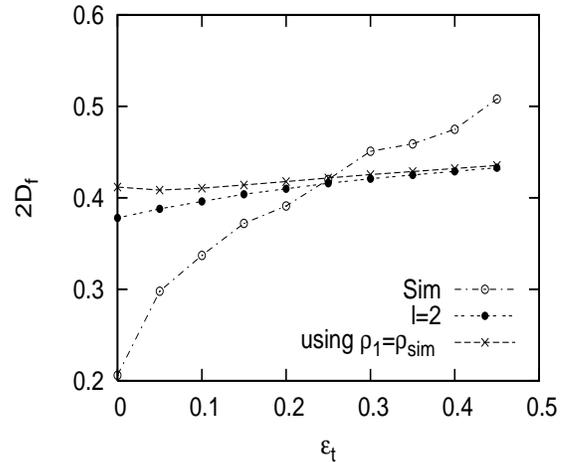}
    \caption{Comparison of the front diffusion coefficient obtained
    analytically (from Eq. (\ref{e8}) and using $\rho_1$ corresponding
    to $l=2$) with
    the simulation results for different values of $\epsilon_t$ while
    keeping $D=0.25$, $\epsilon$=0.025 fixed. We note that when 
    $\epsilon_t$=$D=0.25$, the simulation result matches with the 
    analytic one.}
    \label{fig:diff_coeff}
    \end{figure}

   In Fig. (\ref{fig:diff_coeff}), we have shown the simulation
   results of the front diffusion coefficient and compared it with
   the results obtained by using the mean field and simulation value of 
   $\rho_1$ in the equation(\ref{e8}). Here, we notice the 
   following interesting features: (1) when $D=\epsilon_t$, the 
   analytical value $D_f^{ana}$ matches well with the simulation 
   result $D_f^{sim}$. (2) when $\epsilon_t > D$, the $D_{f}^{sim}
   > D_{f}^{ana}$  (3) when $\epsilon_{t}< D $, $D_{f}^{sim}<
   D_{f}^{ana}$. The origin of the above discrepancy between the 
   simulation and analytical results can be traced to the master
   equation (\ref{e4}), where we have neglected the temporal velocity
   correlations. The expression for the asymptotic front diffusion 
   coefficient with temporal correlations in velocity is given as: 
   \begin{eqnarray}{\label{e20}}
   D_f=D_0+\sum_{t=1}^{\infty}C(t)    
   \end{eqnarray}
   where, $D_0$ is the front diffusion coefficient by neglecting correlations
   as given by Eq. (\ref{e8}) and $C(t)$ is the 
   temporal velocity correlation defined through,
   \begin{eqnarray}{\label{e21}}
   C(t)=<v(t')v(t'+t)> - <v(t')><v(t'+t)>,\nonumber\\
   \end{eqnarray}
   where, $v(t)$ is the displacement of the front at time $t$.   
    \begin{figure}
     \centering
    \includegraphics[bb=50 49 260 210]{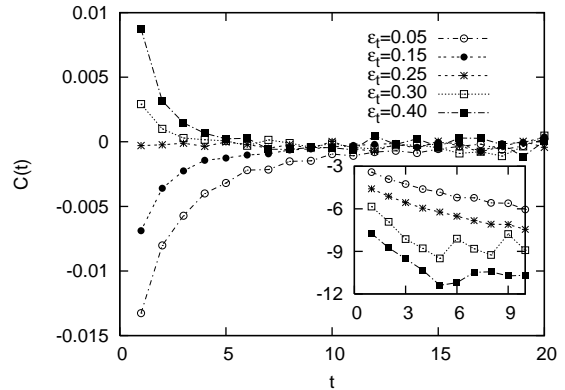}
    \caption{Simulation results for velocity correlation of the 
    front with time $t$, for different values of $\epsilon_t$ while keeping 
    $D=0.25$, $\epsilon$=0.025 fixed. We notice that when
    $\epsilon_t=0.25$, this correlation is zero. Inset: log-normal plot
    for $C(t)$ versus $t$ with $D=0.05, 0.10, 0.30, 0.40$ from top to bottom.}
    \label{fig:vel_corr}
   \end{figure}
   In Fig. (\ref{fig:vel_corr}), we have plotted the temporal velocity
   correlation $C(t)$ for different values of $\epsilon_t$. For  
   $\epsilon_t>D$, we observe positive correlation while for 
   $\epsilon_t<D$, it is negative and for $\epsilon_t=D$, $C(t)$ seems 
   to vanish for all $t$. Thus, $\epsilon_t=D$ is a special case, where the
   front particle moves like an uncorrelated random walker. 
     
    In the following, we explicitly show that for the special case $\epsilon_t=D$,
    two consecutive steps of the leading particle are uncorrelated, i.e., $C(1)=0$
    in the steady state. Since at most two sites behind the front can be affected
    in two consecutive steps, we consider 4 states corresponding to $l=2$, namely, 
    $\{001,011,101,111\}$ with the rightmost $'1'$ representing the front. In order
    to find $C(1)=<v(t)v(t+1)>-<v(t)><v(t+1)>$, we write $<v(t)v(t+1)>=R_{++}-
    R_{+-}-R_{-+}+R_{--}$, where $R_{ij}$ denotes the 'flux' $R^{001}_{ij}+
    R^{011}_{ij}+R^{101}_{ij}+R^{111}_{ij}$ for taking two consecutive steps as 
    $i=+/-$ and $j=+/-$.
    Here, for example, $R^{001}_{--}$ is the flux of two consecutive negative
    steps starting from the state $001$. The only way it can occur is if the
    front particle takes two diffusive moves to the left and thus, $R^{001}_{--}=
    D^2P_{001}$, where $P_{001}$ is the steady state weight of the configuration
    $001$. Considering all such two successive moves in the each state, we write 
    the following expression for  $R_{++},R_{+-},R_{-+}$ and 
    $R_{--}$,
   \begin{eqnarray}{\label{e22}}
   R_{++}&=&D^2+2\epsilon D+\epsilon_t D+\epsilon^2+(\epsilon_t D
         +\epsilon_t\epsilon)\{P_{001}+P_{101}\},\nonumber\\
   R_{+-}&=&D^2,\nonumber\\
   R_{-+}&=&(D^2+D\epsilon+D\epsilon_t)P_{001}+(D^2+D\epsilon)P_{101}
            ,\nonumber\\
   R_{--}&=&D^2P_{001}.
   \end{eqnarray}
   Now using Eqs.(\ref{e22}), we have
   \begin{eqnarray}{\label{e23}}
   <v(t)v(t+1)>&=&\epsilon^2+D\epsilon_t+2\epsilon D
     +\{\epsilon_tD+\epsilon_t\epsilon\nonumber\\-D^2-D\epsilon)\}
       P_{101}&+&\{\epsilon_t\epsilon-D\epsilon\}P_{001},
   \end{eqnarray}
   and similarly,
   \begin{eqnarray}{\label{e24}}
   <v(t)>=<v(t+1)>=\epsilon+\epsilon_t-\{P_{011}+P_{111}\}
             (\epsilon_t-D)\nonumber.\\
   \end{eqnarray}
    Using Eqs.(\ref{e23}) and (\ref{e24}) we can find the value of 
    $<v(t)v(t+1)>$ or $<v(t)>$ by finding the 
    probabilities of different states, which is harder to compute exactly.
    However, when $\epsilon_t=D$, we find that $<v(t)v(t+1)>-<v(t)><v(t+1)>$
    is independent of all the probabilities and is equal to zero, 
    i. e., two successive steps are uncorrelated as observed in the simulation
    result (Fig.\ref{fig:vel_corr}). We also note that for this special
    case the above analysis does not involve any approximation i.e. it
    is exact. We also notice that when $\epsilon_t\ne D$, $<v(t)v(t+1)>\ne
    <v(t)><v(t+1)>$, i.e.,  the front motion is correlated. Preliminary
    fits suggest that the the temporal velocity correlation has the form,
    $C(t)\sim t^\alpha e^{-\beta t}$.

  \section{Conclusion}
   In the present paper, we studied the reaction diffusion system 
   $A\rightarrow 2A, A\rightarrow3A$ in one dimension. Treating the 
   rightmost occupied site as a front
   we compute the front velocity analytically using different
   approximate methods. In fixed site representation one can systematically
   improve upon
   the estimate by studying the evolution of particles at larger number
   of sites behind the front. The results from two particle representation 
   and mixed representation show excellent agreement with 
   the simulation results. We also observed that the velocity depends
   on $\epsilon_t$ as a power law. As far as the computation of front
   diffusion coefficient is concerned, we notice that one needs to take
   into account the temporal velocity correlation. In fact the observed
   temporal correlations in the front dynamics changes sign with sign of
   $\epsilon_t-D$. For $\epsilon_t>D$ or $\epsilon_t<D$, front
   moves like a positively or negatively correlated random walk 
   and for $\epsilon_t=D$, the temporal correlations in steps vanish
   and the front particle moves like a simple uncorrelated random walker.
   An interesting generalization of the process would be to include the
   annihilation of particles as well i.e., $2A\rightarrow A$ with rate $W$.
   The case $\epsilon_t=0$ and $W=D$ is exactly solvable \cite{bA} and 
   both temporal and spatial correlations vanish. For non-zero $D$,
   $\epsilon_t$ and $W$, simulations show that temporal correlations 
   vanish on the plane $D=\epsilon_t+W$, although spatial correlations do not.
   Computation of the previous section can be extended to this case
   to show that consecutive steps are temporally uncorrelated. A general
   proof to show that $C(t)$ vanishes for all $t$ remains an interesting
   open problem.
              
  \end{document}